\documentclass[10pt,a4paper,twoside,pdf]{article}
\usepackage{geometry} 
\usepackage{graphicx}
\newcommand{\gae}
{\lower 2pt \hbox{$\, \buildrel {\scriptstyle >}\over {\scriptstyle \sim}\,$}}
\newcommand{\lae}
{\lower 2pt \hbox{$\, \buildrel {\scriptstyle <}\over {\scriptstyle \sim}\,$}}

\title{Resonant Behavior of Comet Halley and the Orionid Stream}
\author
{A. Sekhar$^{1,2}$ and D. J. Asher$^1$ \\
\\ $^1$Armagh Observatory, College Hill, Armagh BT61\ 9DG, United Kingdom \\
\\ $^2$Queen's University of Belfast, University Road, Belfast BT7 1NN , United Kingdom\\ \\ {E-mail: asw@arm.ac.uk , asekhar01@qub.ac.uk}}

\date{{\bf Received}: 24 Aug 2012; {\bf Accepted}: 6 Mar 2013; {\bf Meteoritics \& Planetary Science}} 

\begin{document}

\maketitle{{\bf Abstract}- Comet 1P/Halley has the unique distinction of having a
very comprehensive set of observational records for almost every perihelion
passage from 240 B.C. This has helped to constrain theoretical models
pertaining to its orbital evolution. Many previous works have shown the
active role of mean motion resonances in the evolution of various meteoroid
streams. Here we look at how various resonances, especially the 1:6 and 2:13
mean motion resonances with Jupiter, affect comet 1P/Halley and thereby enhance the
chances of meteoroid particles getting trapped in resonance, leading to
meteor outbursts in some particular years.  Comet Halley itself librated in
the 2:13 resonance from 240 B.C. to 1700 A.D. and in the 1:6 resonance from
1404 B.C. to 690 B.C., while stream particles can survive for timescales of
the order of 10,000 years and 1,000 years in the 1:6 and 2:13 resonances
respectively.  This determines the long term dynamical evolution and stream
structure, influencing the occurrence of Orionid outbursts.  Specifically we
are able to correlate the occurrence of enhanced meteor phenomena seen
between 1436-1440, 1933-1938 \& 2006-2010 with the 1:6 resonance and meteor
outbursts in 1916 \& 1993 with the 2:13 resonance.  Ancient as well as modern
observational records agree with these theoretical simulations to a very good
degree.

{ \bf Keywords}: comet, meteor, orbit }

\section{Introduction}

Various contributions from ancient civilizations have helped in making a
detailed observational record (Yeomans \& Kiang 1981) of comet 1P/Halley for
almost every perihelion passage right from 240 B.C\@.  There are no credible
observations relating to this comet before 240 B.C\@.  Furthermore comet
Halley has reliably determined (Yeomans \& Kiang 1981) perihelion passage
times (the first calculations done by Halley 1705 using Newton 1687) and
orbital elements back till 1404 B.C., beyond which the uncertainty in the
orbit starts to increase because of a significant close encounter with Earth
at a distance of about 0.04 AU.

Historical confirmations of the annual nature of the Orionid meteor shower
date back to as early as Edward Herrick's observations in 1839 (Lindblad \&
Porubcan 1999) and Alexander Herschel's radiant determination (Denning 1899)
in 1864 (Herschel 1866). Many ancient records of meteors seen in October from the Chinese,
Japanese and Korean civilisations (Imoto \& Hasegawa 1958, Zhuang 1977) could also
correspond to the Orionid shower. Nevertheless the association of the stream
with comet Halley and explaining the differences of the Orionid shower
compared to the Eta Aquariids (which have the same parent body) has been a
very challenging task (McIntosh \& Hajduk 1983, McIntosh \& Jones 1988) which interested many
theoreticians for decades. Coincidentally it is widely believed that Sir
Edmond Halley was the first (by 1688) to suggest that meteors were of cosmic
origin (Williams 2011).

Comet Halley might lose approximately 0.5\% of its mass during every
perihelion passage (Whipple 1951, Kresak 1987) which would predict its
physical lifetime to be a couple of hundred revolutions or $\sim$15 kyr.  The
dynamical lifetime (time scale to remain on any kind of Halley type comet
orbit i.e.\ orbital period from 20 to 200 years) of 1P/Halley is estimated to
be of the order of 100,000 years (Hughes 1985, Hadjuk 1986, Steel 1987,
Bailey \& Emel'yanenko 1996).  Bailey \& Emel'yanenko (1996) showed that
Halley undergoes Kozai resonance (Kozai 1979) during its long term evolution.
It is reasonable to believe (see also Section 4 below) that Halley has been
on an orbit comparable to its present one (with perihelion distance $q$ \lae
1 AU), and outgassing particles thereby populating the Orionid stream, for a
couple of tens of kyr.

It is interesting to note that the zenithal hourly rates (ZHR) of Orionids
are non-uniform (Miskotte 1993, Rendtel \& Betlem 1993, Rendtel 2007, Trigo-Rodriguez et al.\ 2007, Arlt et al.\ 2008, Rendtel 2008,  Kero et al.\ 2011, IMO
database) with respect to each
year. Many previous works have discussed the active role of mean motion
resonances (MMR) in the dynamical evolution of various meteoroid streams
(e.g.\ Asher et al.\ 1999, Emel'yanenko 2001, Asher \& Emel'yanenko 2002,
Ryabova 2003, Ryabova 2006, Jenniskens 2006, Vaubaillon et al.\ 2006,
Jenniskens et al.\ 2007, Christou et al.\ 2008, Soja et al.\ 2011), and
consequent year to year variations in shower activity.  Our work aims to
study the long term evolution of Halley and its associated stream focusing
especially on past resonant behavior.  We model particles ejected from the
comet and try to correlate these with ancient as well as present
observational records of meteor showers.

\section{Resonant Motion of Comet Halley}

Over the time frame during which 1P/Halley's orbit is reliably known,
i.e.\ since 1404 B.C., our calculations show that the comet was resonant in
the past: it was trapped in the 1:6 and 2:13 MMR with Jupiter from 1404
B.C. to 690 B.C. and 240 B.C. to 1700 A.D. respectively.  Integrations were
repeated for different values of non-gravitational parameters (Marsden et al.\ 1973, Marsden \&
Williams 2008) to ensure that this resonant pattern is not sensitive to small
changes in non-gravitational forces.  Fig.\ 1 shows the 1:6 resonant argument
librating from 1404 B.C. to about 690 B.C., and Fig.\ 2 shows the 2:13
resonant argument librating from 240 B.C. to 1700 A.D.

\begin{figure}[h] 
   \centering
   \includegraphics[width=4in]{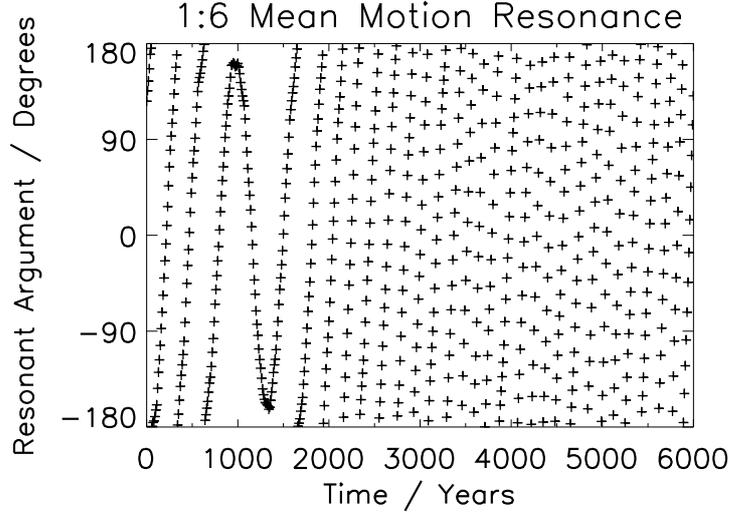} 
   \caption{Evolution of 1:6 resonant argument of 1P/Halley over 6000 years
     from 2404 B.C.}
   \label{X}
\end{figure}

All the orbit integrations in this work were done using the MERCURY package
(Chambers 1999) incorporating the RADAU algorithm (Everhart 1985), and
including the sun and eight planets, whose orbital elements were taken from
JPL Horizons (Giorgini et al.\ 1996).  Elements for the comet were taken from
Yeomans \& Kiang (1981).

\begin{figure}[h] 
   \centering
   \includegraphics[width=4in]{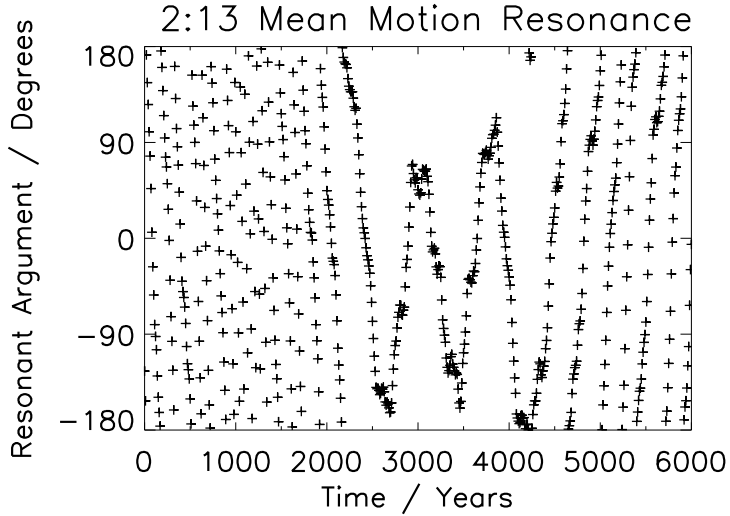} 
   \caption{Evolution of 2:13 resonant argument of 1P/Halley over 6000 years
     from 2404 B.C.}
   \label{X}
\end{figure}

Since the comet has a retrograde orbit, the change in the definition of
longitude of pericentre $\varpi$ (Saha \& Tremaine 1993, Whipple \& Shelus
1993) was incorporated while computing the resonant arguments:

\begin{equation}
\varpi=\Omega-\omega
\end{equation}
where $\Omega$ and $\omega$ are longitude of ascending node and argument of
pericentre respectively.

\begin{figure}[h] 
   \centering
   \includegraphics[width=4in]{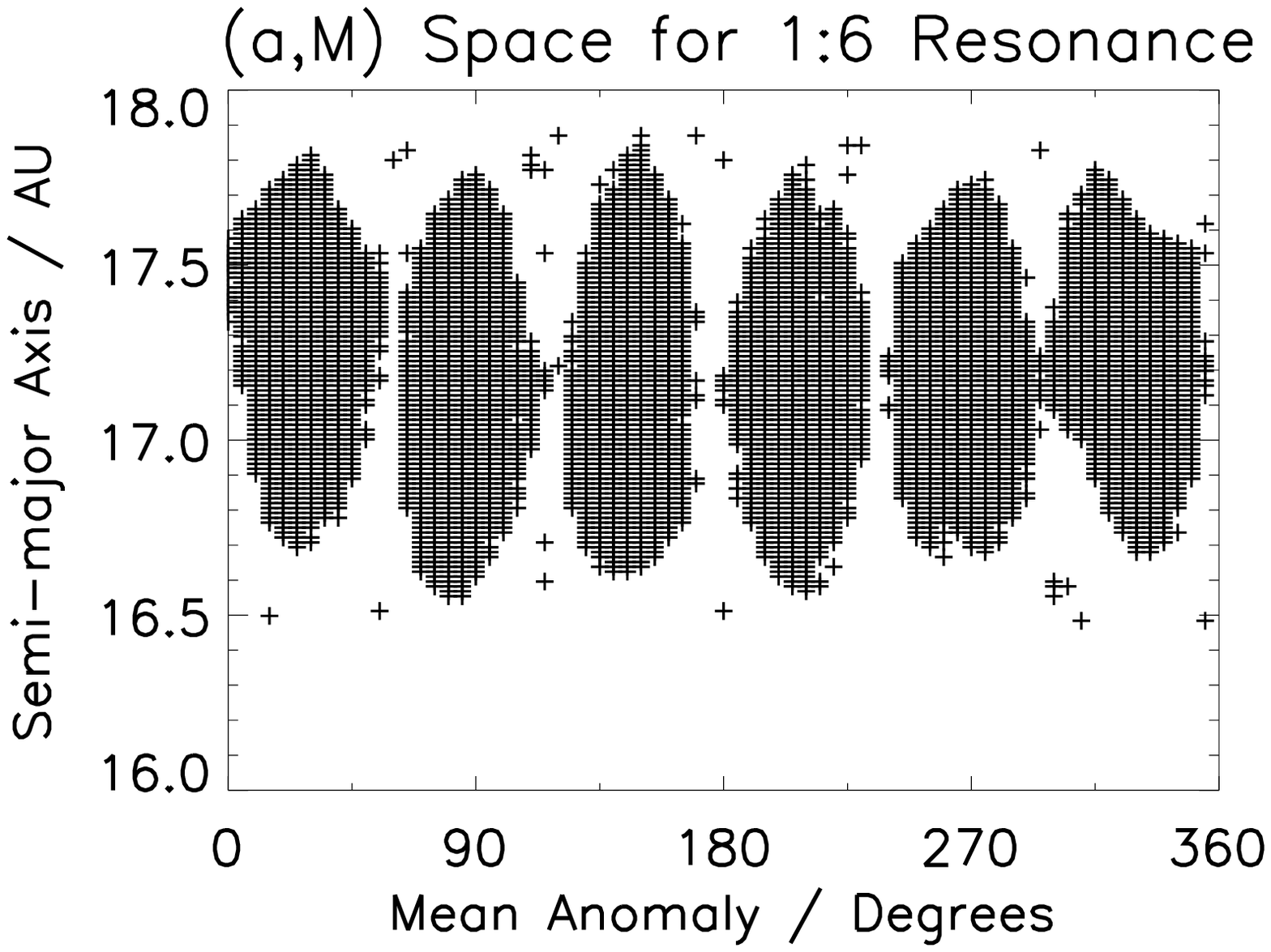} 
   \caption{(a,M) space for 1:6 resonance showing regions where particles
undergo resonant librations, as a function of initial semi-major axis and
mean anomaly at the initial epoch JD 1208880.5.}
   \label{X}
\end{figure}

In order to absolutely confirm the librating versus circulating behavior of
the resonant argument during the time frames mentioned above, various
combinations of terms to define the resonant argument (Murray \& Dermott
1999, Sections 6.7 and 8.2) according to the D'Alembert rules were verified.
Mathematically the D'Alembert rule is given by Equation 2, and Equations
3 and 4 should be satisfied for Equation 2 to be valid.

In the case of the p:(p+q) mean motion resonance
\begin{equation}
\sigma=p\lambda_j-(p+q)\lambda_c+k_1\varpi_c+k_2\varpi_j+k_3\Omega_c+k_4\Omega_j
\end{equation}
where q is the order of resonance, $\sigma$ and $\lambda$ denote resonant
argument and mean longitude respectively, and subscripts c and j stand for
the comet and Jupiter.

\begin{equation}
k_1+k_2+k_3+k_4=q
\end{equation}

\begin{equation}
k_3+k_4= 0, 2, 4, ......
\end{equation}

For the 1:6 MMR (q=5) there are 28 combinations and each of them was checked,
verifying the interval 1404 to 690 B.C. shown in Fig.\ 1.  For the 2:13 MMR
(q=11) there are 182 combinations of which 50 were checked, all of them
verifying the result of Fig.\ 2.  In Figures 1 and 2, $\sigma$ is plotted for
the combinations shown in Equation 5 and 6 respectively.

\begin{equation}k_1=5, k_2=k_3=k_4=0\end{equation}
\begin{equation}k_1=5, k_2=4, k_3=1,k_4=1\end{equation}

When the comet itself is resonant, there are more chances for the ejected
meteoroid particles to get trapped in resonance which in turn would enhance
the chances for meteor outbursts in future years. That is an important
motivation for looking into the resonant behavior of the parent body.

\begin{figure}[h] 
   \centering
   \includegraphics[width=4in]{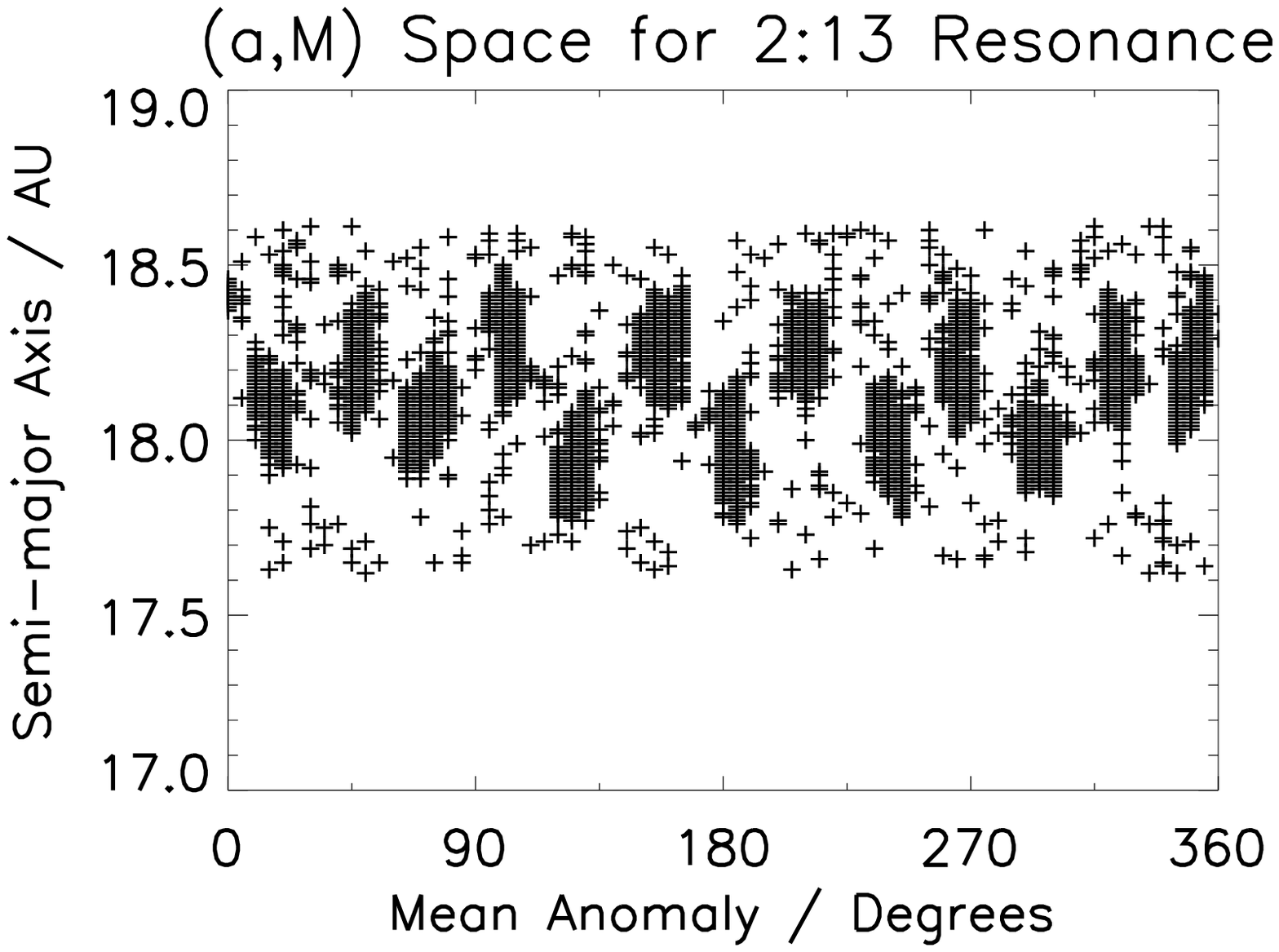} 
   \caption{(a,M) space for 2:13 resonance (cf. Fig. 3); initial epoch JD 1633920.5}
   \label{X}
\end{figure}

\section{Resonant Structures in the Orionid Stream}

\subsection{General Schematic}

Figures 3 and 4 shows the general schematic of the geometry of resonant zones
in the case of 1:6 ($a_n$=17.17 AU) and 2:13 ($a_n$=18.11 AU) resonances
respectively.  Here we quote $a$ = $a_n$ = the `nominal resonance location'
(Murray \& Dermott 1999, Section 8.4), which is the value of semi-major axis
corresponding to a resonant orbital period assumes the most simple case where
($d\varpi/dt$) (which denotes orbital precession) is zero, i.e.  as implied by Kepler's third law (Kepler 1609, 1619)

\begin{figure}[h] 
   \centering
   \includegraphics[width=4in]{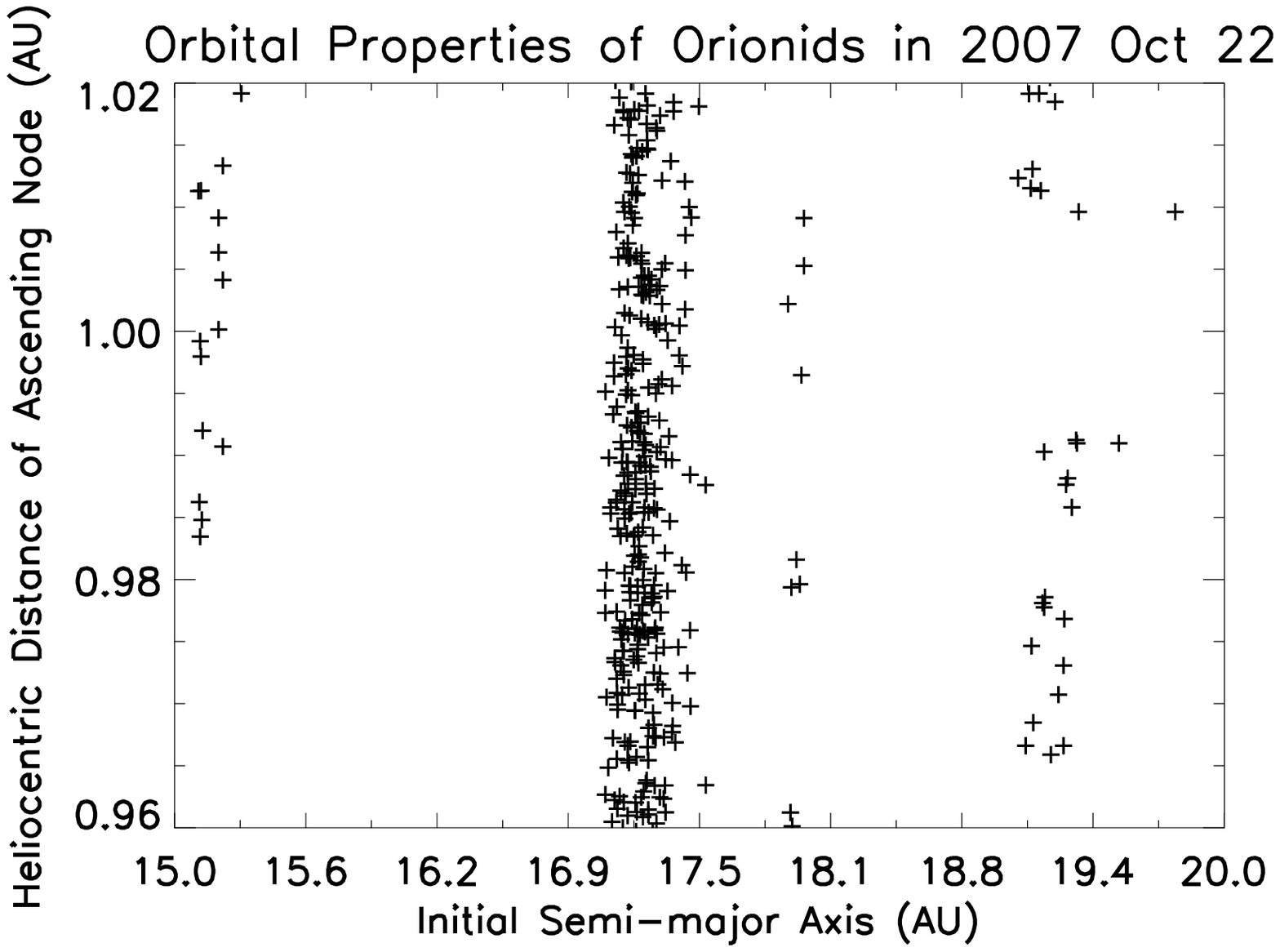} 
   \caption{Ascending Nodal Distance in 2007 vs Initial Semi-major Axis of Meteoroids in -910.}
   \label{X}
\end{figure}

In a real case ($d\varpi/dt$) is never exactly zero, e.g.\ with the 1:6
resonance (resonant argument as per Equation 5):

 \begin{equation}
\sigma = \lambda_j-6\lambda_c+5\varpi_c
\end{equation}

If we assume, as a time average, for resonant libration:
 \begin{equation}
d\sigma/dt = 0
\end{equation}

then
\begin{equation}
\lambda_j-6(\varpi_c+M_c)+5\varpi_c = \sigma
\end{equation}

where M denotes the mean anomaly.\\

Simplifying the expression we get
\begin{equation}
\lambda_j-\varpi_c-6M_c = \sigma
\end{equation}

Differentiating Equation 10 on both sides with respect to time and using Equation 8,

\begin{equation}
(d\lambda_j/dt)-(d\varpi_c/dt)-6(dM_c/dt) = 0
\end{equation}

From our numerical integrations we find that (d$\varpi_c$/dt) is always
positive for these resonant particles. If (d$\varpi_c$/dt) is positive, then
we require the rate of change of mean anomaly to be smaller compared to the
`nominal' case, which in turn means an increase in the value of semi-major
axis.

\begin{equation}
a_{real}=a_n+\Delta a
\end{equation}

Therefore the actual resonant value of semi-major axis would be slightly
greater than the ones given in this section. A much more comprehensive and
general description about this subject and its application to meteor streams
can be found in Emel'yanenko (2001).

\begin{figure}[h] 
   \centering
   \includegraphics[width=4in]{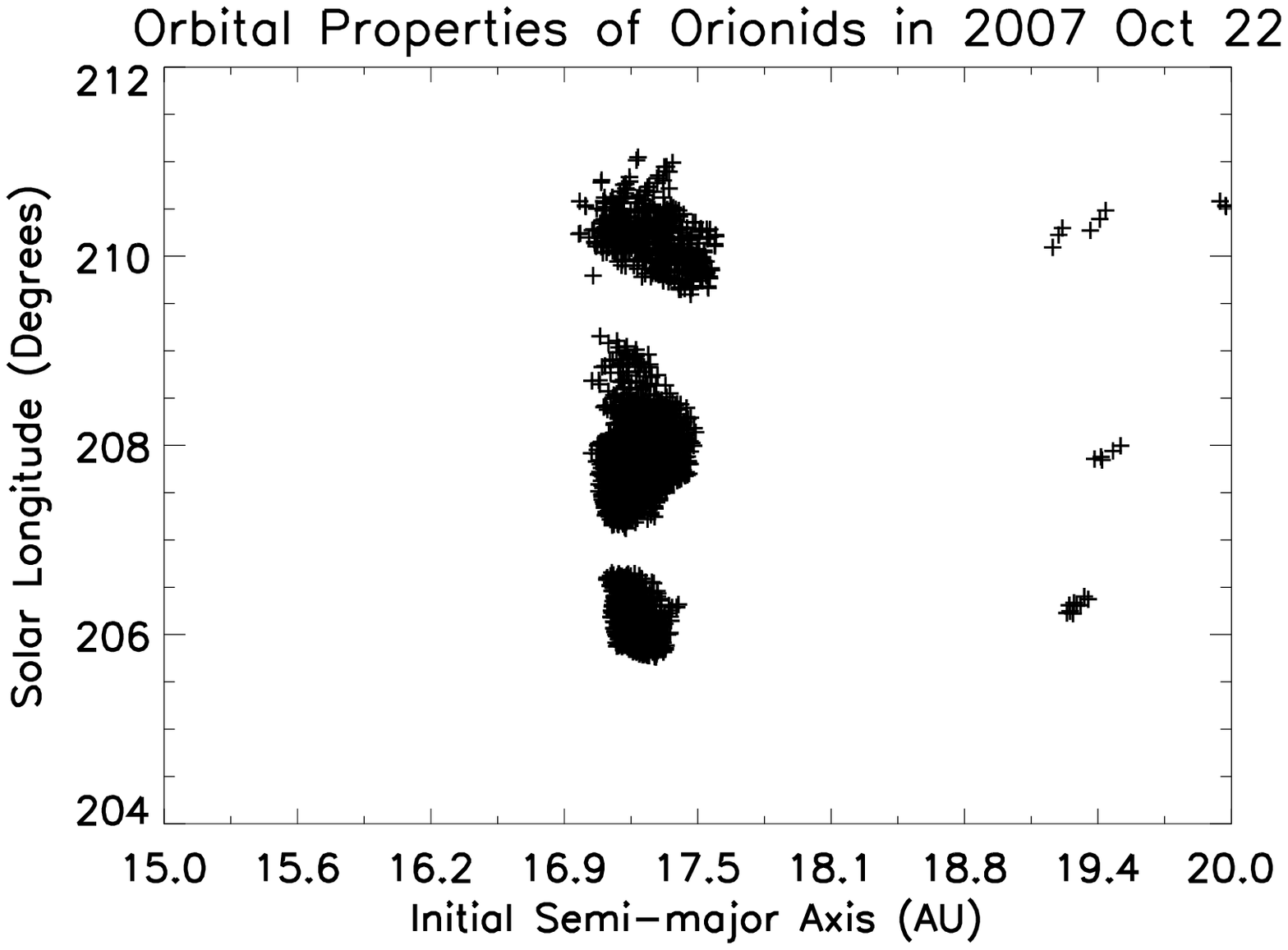} 
   \caption{Solar Longitude in 2007 vs Initial Semi-major Axis of Meteoroids in -910.}
   \label{X}
\end{figure}

Integrations were done by taking 7200 particles, varying the initial $a$ from
16.5 to 17.9 in steps of 0.014 AU for 1:6 MMR and from 17.6 AU to 18.6 AU in
steps of 0.01 AU for 2:13 MMR, and initial M from 0 to 360 degrees in steps
of 5 degrees, keeping all other orbital elements (namely e, i, $\omega$ and
$\Omega$) constant. The starting epochs for Fig 3 and Fig 4 are 1P/Halley's
perihelion return times in 1404 B.C. and 240 B.C. respectively. All the
particles were integrated for 2,000 years using the RADAU algorithm with
accuracy parameter set to $10^{-12}$. Output data were generated for every 10
years. Resonant particles were identified by employing a simple technique which
looks at the overall range of the resonant arguments (for various combinations allowed by D'Alembert rules, see Section 2) of all particles during the whole integration time to check when there is no circulation. Checks on an extensive set of representative particles covering many different libration amplitudes confirm that those particles filtered by this algorithm will have librating resonant arguments which thereby confirm the presence of respective mean motion resonances.

\begin{figure}[h] 
   \centering
   \includegraphics[width=4in]{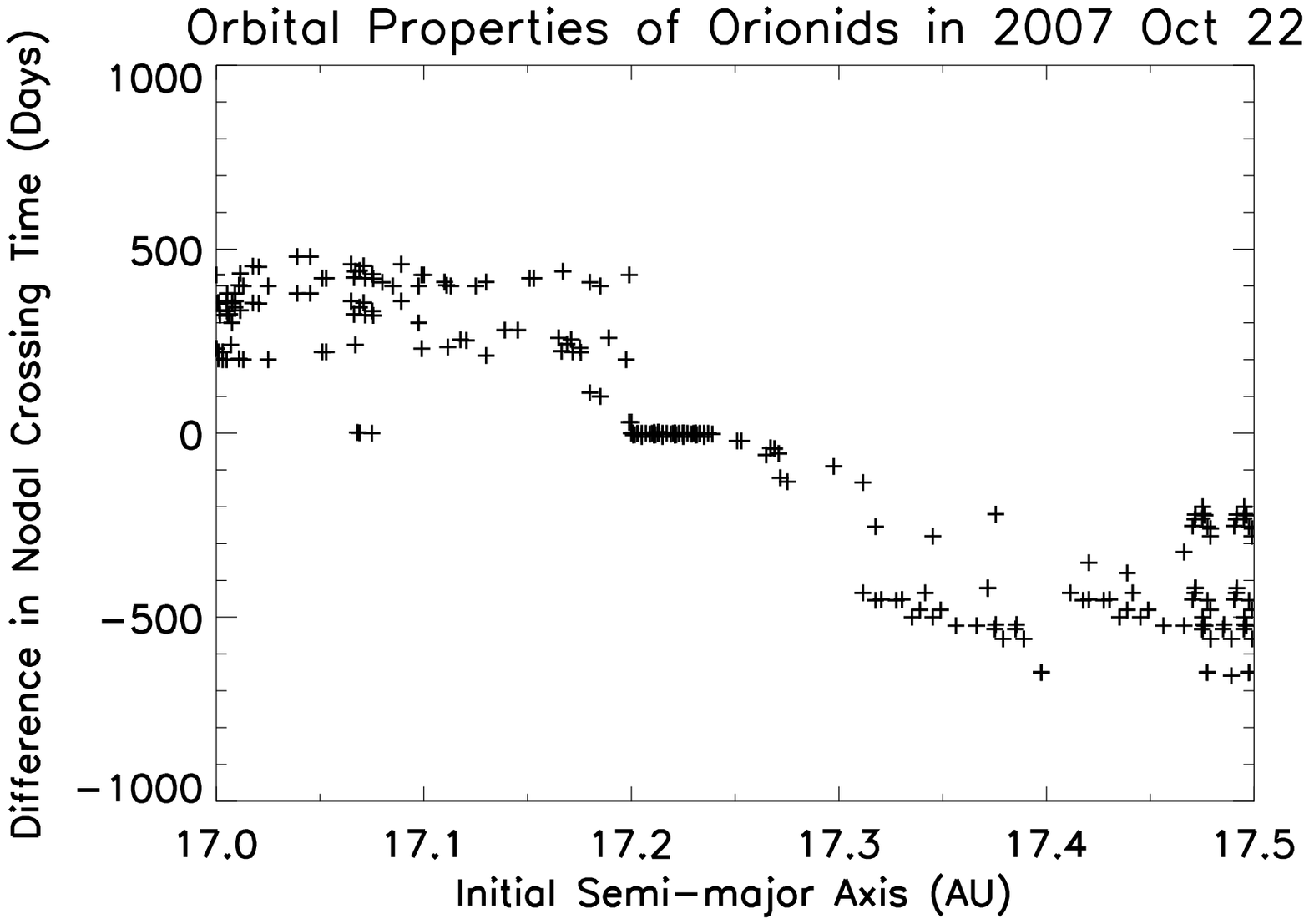} 
   \caption{Difference in Nodal Passage Times in 2007 vs Initial Semi-major Axis of Meteoroids in -910.}
   \label{X}
\end{figure}

Figures 3 and 4 give a general picture of these resonances: we can visualize
6 or 13 resonant zones spaced in mean anomaly along the whole orbit, each
zone consisting of individual librating particles (cf.\ Emel'yanenko
1988). These zones, or clouds of resonant particles, are preserved for as
long as substantial numbers of particles continue to librate.  Our test
integrations showed that for some particular ejection epochs the 1:6 MMR is
exceptionally effective in retaining the compact dust trail structures for as
long as 30,000 years. However particles disperse in mean anomaly much faster
(in a few thousands of years) in the case of the 2:13 resonance and do not
show such high stability. Typically the rule of thumb is that the higher the
order of resonance (denoted by q, see equation 3), the lower the strength of
the resonance.

\begin{figure}[h] 
   \centering
   \includegraphics[width=4in]{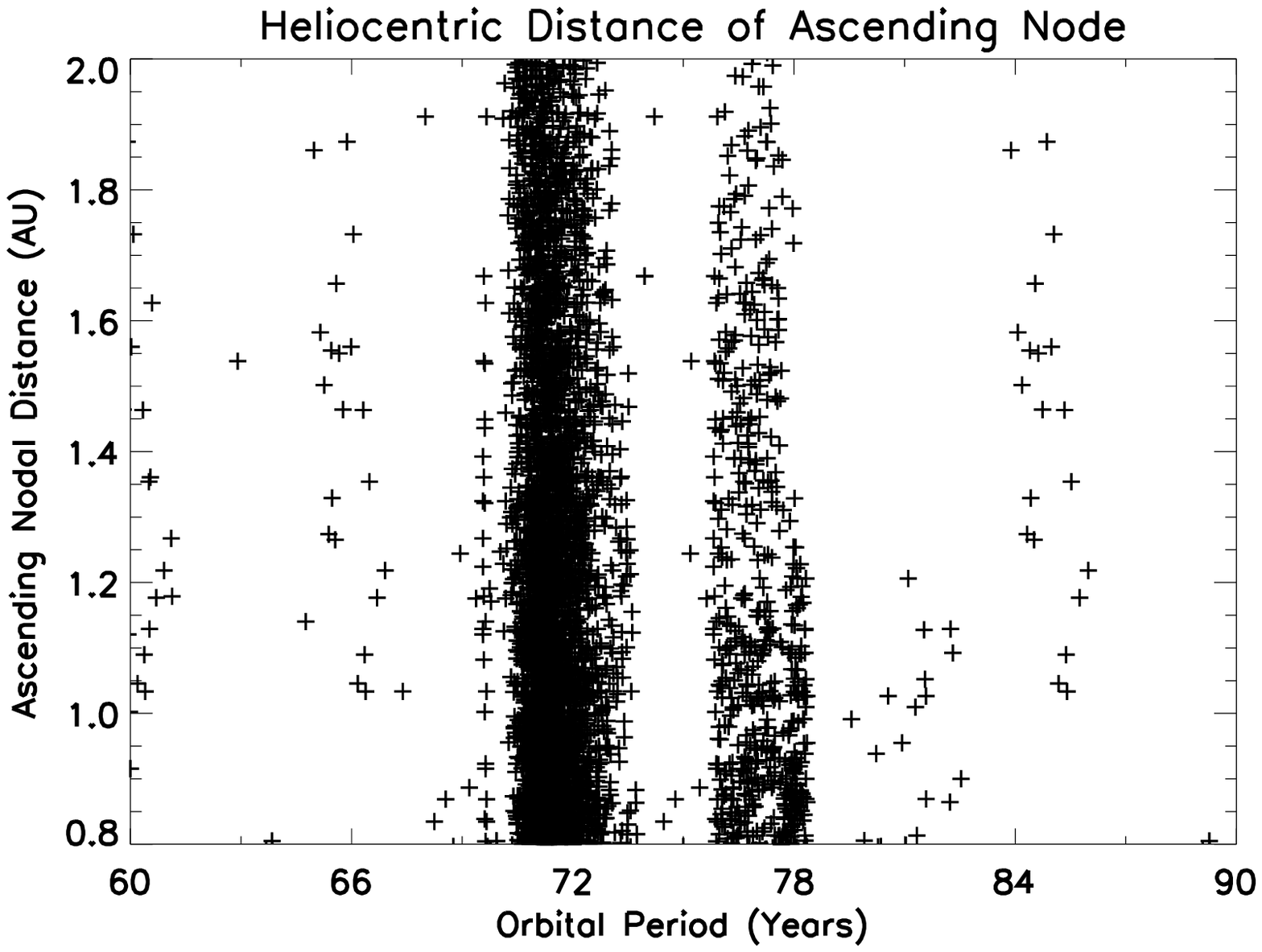} 
   \caption{Ascending Nodal Distance in 2007 vs Initial Orbital Period of Meteoroids in -910.}
   \label{X}
\end{figure}

A necessary condition for a resonant meteor outburst is that the Earth should
encounter one of these clusters of resonant particles, i.e.\ when the Earth
misses these clusters, there is no enhancement (which is the common case in
most years) in meteor activity, at least due to the MMR mechanism.  Of course
there are various other factors like nodal distance, solar longitude, date
and time of intersection of the meteoroid with Earth, geocentric velocity
etc.\ which play a key role in confirming the occurrence and characteristics
of a meteor outburst or storm (see Section 3.2).  It is possible in reality
firstly that many resonant zones would only be partially filled (unlike the
uniform pattern shown in Figures 3 and 4), and secondly that within a given
resonant zone there is significant fine structure which could lead to
enhanced meteor phenomena if Earth happens to pass exactly through the
densest parts.

Similar plots to Figures 3 and 4 reveal mean anomaly distributions of
resonant particles in the long term (a few millennia).  The cluster of
particles in a single 1:6 resonant zone occupies a much longer part of the
orbit (covering 5-6 years) than the equivalent for a 2:13 resonant zone (only
1-2 years).  This means that for the 1:6 MMR there can be 5-6 consecutive
years of enhanced meteor activity (depending on the exact parts of the
resonant stream's fine structure encountered by the Earth), compared to just
1-2 years of outburst possibilities from 2:13.  The 1:6 resonance is also
more effective in trapping considerably larger numbers of particles compared
to 2:13. Hence meteor outbursts from the 1:6 resonance would have a higher
intensity than those due to the 2:13 resonance in most cases, though
dependent again on the fine structure within the respective resonances. Since
there is a long record of observations for Orionids (Rendtel 2008) the same
pattern could be compared and justified. The maximum ZHR in 2007 was about 80
(Arlt et al.\ 2008) whereas in 1993 it was about 35 (Rendtel \& Betlem 1993);
cf.\ normal rates of 20-25.

\begin{figure}[h] 
   \centering
   \includegraphics[width=4in]{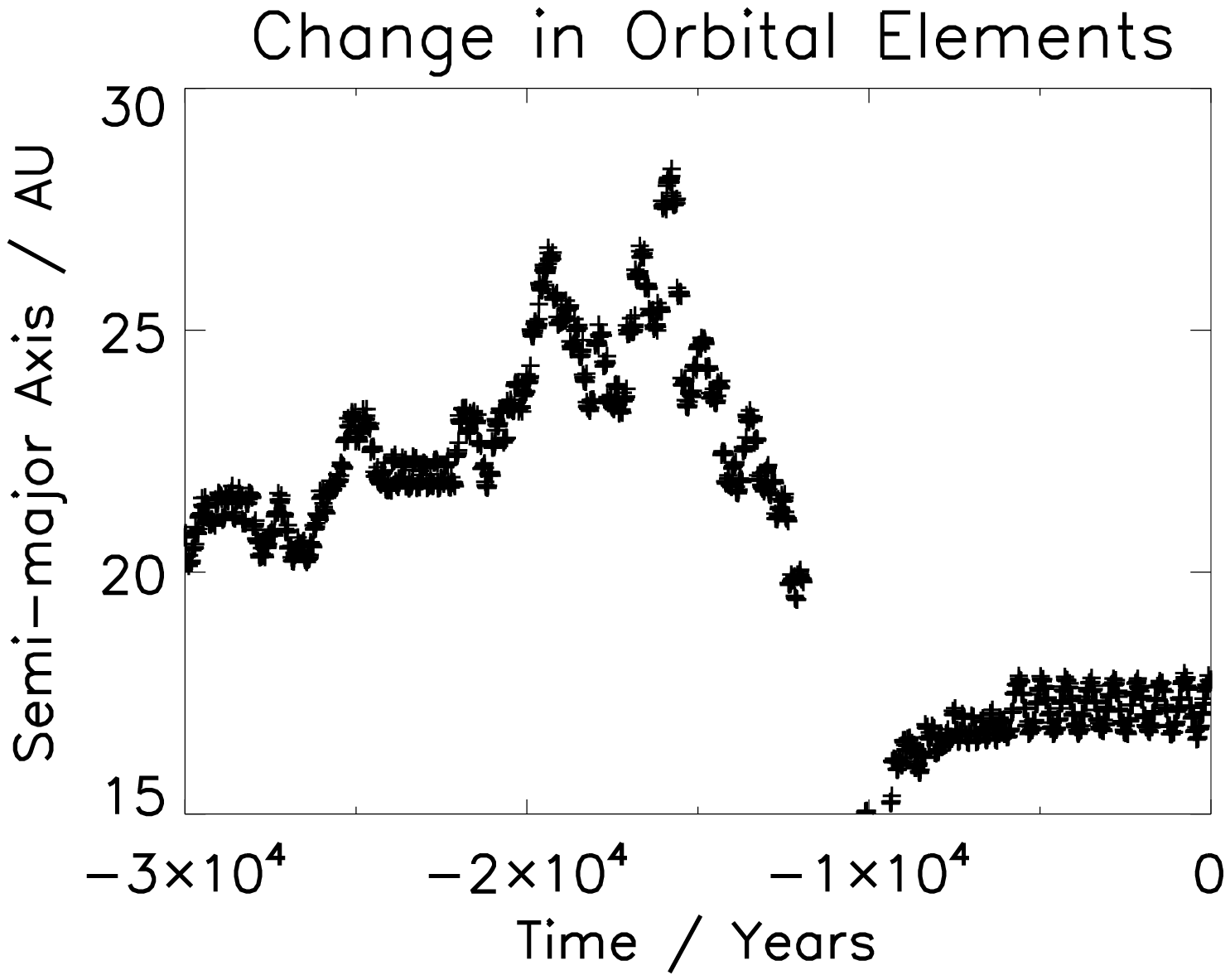} 
   \caption{Evolution of 1P/Halley's semi-major axis in an
     integration going back in time from 240 B.C.}
   \label{X}
\end{figure}

When the comet is resonant, it would remain in a single resonant zone and
populate that particular zone. When the comet goes out of resonance, it would
keep traversing between zones and thereby populate different resonant zones
gradually, implying that meteor outbursts could come from different
zones. One of the main aims of this work is to correlate particular resonant
zones to past and present meteor outbursts. Our calculations show that the
outbursts during 1436-1440, 1933-1938 and 2006-2010 (Section 3.2) are from
the same 1:6 resonant zone, specifically the same one in which the comet
librated from 1404 B.C. to 690 B.C. Also a future meteor outburst in 2070
would occur due to the particles in the same 2:13 resonant zone which caused
increased meteor activity in 1916 and 1993.  Halley presumably released many
meteoroids into the 2:13 resonant zone in which it librated from 240 B.C. to
1700 A.D. but most of these meteoroids do not have the precession rate
required for producing Earth-intersecting orbits at the present time.
Comparison with resonant zones are an excellent way to match observations
with theoretical simulations.

\begin{figure}[h] 
   \centering
   \includegraphics[width=4in]{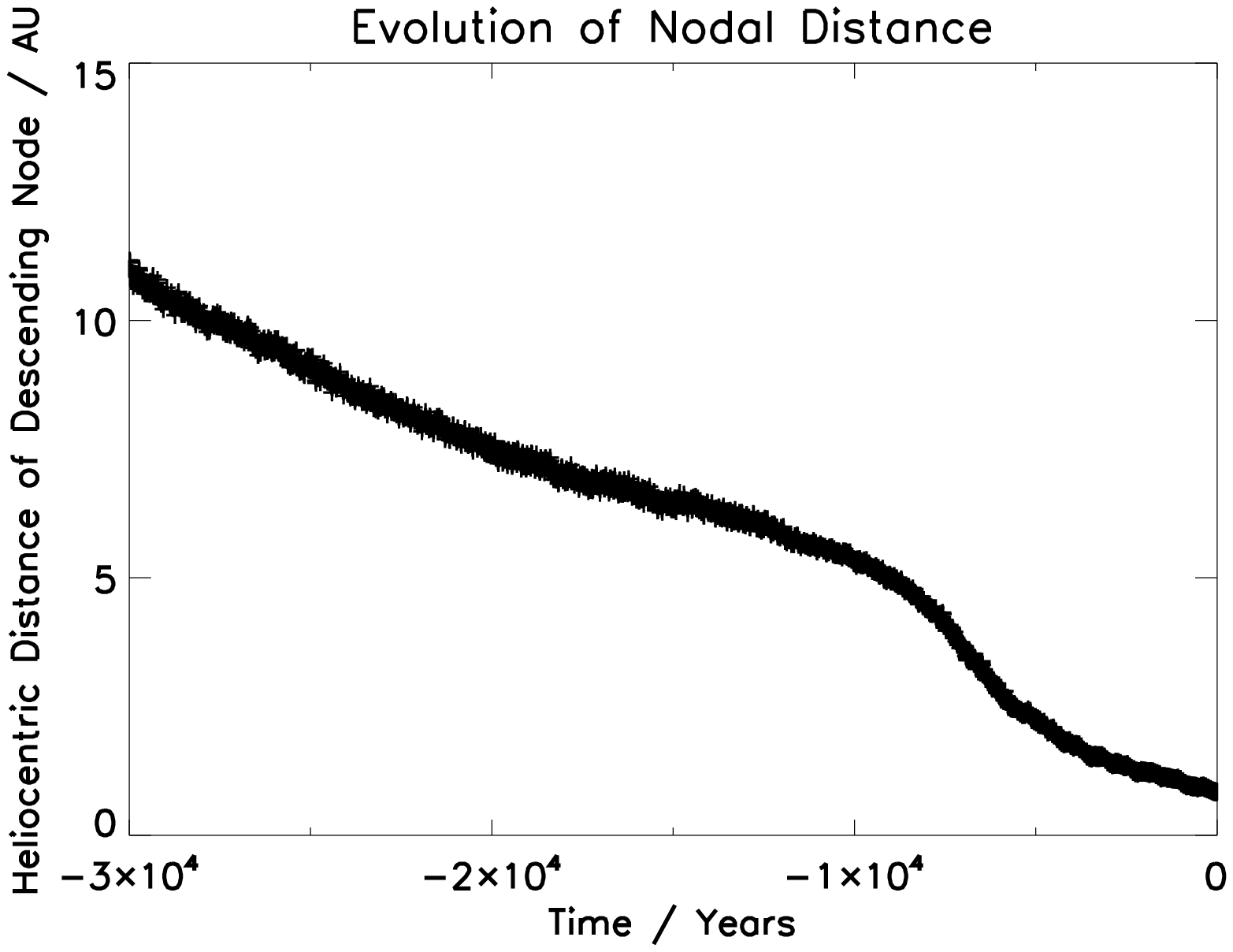} 
   \caption{Heliocentric distance of descending node of 1P/Halley in an
     integration going back in time from 240 B.C.}
   \label{X}
\end{figure}

The numbers of particles trapped in other resonances close to this range of
semi-major axis (16.5 AU to 19 AU) were also checked. For example the
percentage of particles in 1:7 ($a_n$=19.03 AU), 3:17 ($a_n$=16.53 AU), 3:19
($a_n$=17.80 AU) and 3:20 ($a_n$=18.42 AU) resonances with Jupiter are very
low compared to the 1:6 ($a_n$=17.17 AU) and 2:13 ($a_n$=18.11 AU). It should
be pointed out that if particle ejection were centred around the resonant
value of 1:7 MMR, then there would be substantial amounts of resonant
particles which can cause outbursts, but in reality the comet is never near
this resonant semi-major axis in the time frames which we consider in this
work. Also the ratio of particles trapped in the Saturnian resonances of 2:5
($a_n$=17.56 AU), 3:8 ($a_n$=18.34 AU), 5:12 ($a_n$=17.09 AU), 5:13
($a_n$=18.03 AU), 7:17 ($a_n$=17.23 AU), 7:18 ($a_n$=17.90 AU) and 8:19
($a_n$=16.97 AU) are extremely small owing to the overpowering effect of
Jupiter's gravity. Hence significant measurable enhancements in meteor
activity can be ruled out from these obscure Jovian and Saturnian
resonances. Even if such resonant particles encounter Earth it will be almost
impossible to distinguish them because of the lack of any sizeable increase
in ZHR in any year. Hence it should be understood that just the mere fact of
having some resonant particles intersecting Earth does not mean an increase
from normal meteor rates. The sole criterion depends on how effective that
resonance mechanism is in trapping very large numbers of ejected particles and
subsequently avoiding close encounters with other planets.

\subsection{Specific Calculations}

Past observations (Millman 1936, Lovell 1954, Imoto \& Hasegawa 1958, Rendtel
\& Betlem 1993, Dubietis 2003, Rendtel 2007, Trigo-Rodriguez et al.\ 2007, Arlt et al.\ 2008, Spurny \& 
Shrbeny \ 2008, Kero et al.\ 2011) of Orionids have shown enhanced meteor activity in some particular
years. Previous interesting works (Rendtel 2007, Sato \& Watanabe 2007) have
highlighted the significance of 1:6 MMR in explaining the outburst in
2006. According to Sato \& Watanabe (2007), the meteor outburst in 2006 was
caused by 1:6 resonant particles ejected from the comet in -1265 (1266 B.C.),
-1197 (1198 B.C.) and -910 (911 B.C.). All these ejection years can be
directly linked to the time frame in which the comet itself was 1:6 resonant
(Section 2). Hence more meteoroids became trapped in this resonance during
this time frame compared to other years when the comet was not resonant.

Calculations were done on similar lines to Sato \& Watanabe
(2007).  Ejection epochs were set between 1404 B.C. and 1986 A.D.  All the
ejections were done by keeping the perihelion distance and other elements as
constant and by varying the semi-major axis and eccentricity.  In this simple
model, ejection was done at perihelion (M=0) in the tangential direction.
Ejection velocities were set in the range -50 to +50 m/s, i.e.\ both behind
and ahead of the comet, meaning that over all epochs collectively the initial
orbital periods range from 60 to 88 years, encompassing all possible 1:6 and
2:13 resonant particles, positive ejection velocity corresponding to larger
periods.  Radiation pressure and Poynting-Robertson effects were not
incorporated in these calculations.  Because they span a range of orbital
periods, test particles ejected tangentially at perihelion and moving only
under gravitational perturbations are able, over the time frames we consider
here, to represent the motion of all meteoroids released over the comet's
perihelion arc with different velocities and subject to different radiation
pressures (Kondrat'eva \& Reznikov 1985, Asher \& Emel'yanenko 2002).

In order to confirm the correlation between theory and observations, it is
vital to match the time (second half of October) when the meteoroids reach
their ascending node, solar longitude $\lambda_\odot$ (approx. 204-210 degrees) at the node
and heliocentric distance of ascending node (by analyzing the difference
$\Delta$r between heliocentric distances of Earth and ascending node of meteoroid; Earth diameter is about 0.0001
AU). These essential parameters from our simulations (Table 1) can be
matched with real observations (listed in past meteor records).

Each entry in Table 1 is a carefully chosen meteoroid which has the average
value out of many candidate particles covering the small range of orbital
periods that favors an outburst in the given year.  For example, Figures 5,
6 and 7 are plots of heliocentric distance of ascending node, solar longitude
and difference in time of nodal crossing of the particles from the time of
observed outburst (all three parameters computed at 2007 Oct 22) versus
initial semi-major axis of meteoroids (ejected at -910 return). Heliocentric
distance of Earth on 2007 Oct 22 was 0.995 AU. These results show that the
conditions for an outburst to occur (from particles ejected in -910) at the
observed time in 2007 is satisfied if initial semi-major axis is around 17.22
AU, but the total suitable range in initial semi-major axis and ejection
velocities for meteoroids are 17.20 AU \textless a \textless 17.25 AU and
-16.53 m/s \textless v \textless -15.05 m/s respectively. The plots are
similar for other ejection epoch / outburst year pairs as well.

Our simulations indicate meteor outbursts from 1436-1440 A.D. due to 1:6
resonant meteoroids which were ejected around Halley's -1265, -1197, -985,
-910 and -836 returns (details in Table 1). The initial orbital periods of
ejected meteoroids which lead to all five outbursts show that most of them
had almost 6 times the Jovian period. There are also historical observational
records which show heightened activity, indicating hundreds of bright
meteors, in 1436 and 1439 (Imoto \& Hasegawa 1958) and match our
theoretical simulations well. We converted the dates from Julian calendar to
proleptic Gregorian calendar in order that all dates in Table 1 are referred
to a single calendar. No observational records could be traced or identified
for 1437, 1438 and 1440 though. Either there were no observations done in
those years (unfavorable lunar phase is a possible explanation only in
1438) or the meteor outbursts would have been insignificant in 1437,
1438 and 1440 compared to the ones in 1436 and 1439. In our simulations we
find that resonant meteoroids ejected with positive ejection velocity (higher
orbital period) encountered Earth in 1436 and 1439. The ones with negative
ejection velocity (smaller period) encountered Earth in 1437, 1438 and 1440.
Radiation pressure (not included in our integrations) would always act in the
direction which would increase the orbital period of meteoroids, i.e.\
affects the period in the same sense as positive ejection velocities. In
general we expect the peak of the ejection velocity distribution is close to
zero and so the largest number of particles, if affected by radiation
pressure, is represented by particles having positive ejection velocites in
our gravitational integration model. Radiation pressure is having a
detrimental effect (with regard to causing meteor outbursts) when we
calculate that negative ejection velocities are required to produce a meteor
outburst in a particular year. This can explain why we find this trend in
ejection velocities for resonant meteoroids reaching Earth in 1436 and 1439
which caused meteor outbursts (agreeing with past observations as shown in
Imoto \& Hasegawa 1958) and possibly no (or very low) activity in 1437, 1438
and 1440.

We calculate that the meteor outburst in 1993 was due to 2:13 resonant
meteoroids ejected around Halley's -1333, -985, -910 and -835 returns. The
outburst (Miskotte 1993, Rendtel \& Betlem 1993) occurred when solar longitude was between
204.7 to 204.9 degrees, a notably different time compared to other known
outbursts. Our theoretical calculations match this unusually early peaking on
1993 Oct 18. Our simulations also indicate a meteor outburst in 1916 from the
2:13 resonance and there is a hint of enhanced meteor rates from past
observations in 1916 (Olivier 1921) compared to the adjacent years of 1915
and 1917. For the future we predict a similar outburst (like in 1993 because
favorable ejection velocities are similar in both cases) from the 2:13
resonance mechanism in 2070.

The ejection epochs for 1:6 resonant meteoroids which caused continuous
enhanced activity in 2007, 2008, 2009 and 2010 (Trigo-Rodriguez et al.\ 2007, Arlt et al.\ 2008, Kero et al.\ 2011, International Meteor Organization
database) are also given in Table 1. These ejection years correspond to the
time when the comet itself was 1:6 resonant. Hence it is obvious that a large
number of meteoroids would have been trapped into this resonance during those
time frames which would clearly indicate the reason for high ZHR apart from
the contribution due to the inherent geometry (see Section 3.1) of these
zones. Our simulations match the observed ranges (207-210 degrees) of solar
longitude and outburst times (Trigo-Rodriguez et al.\ 2007, Arlt et al.\ 2008, Kero et al.\ 2011, IMO
database) for these outburst years very
well. Even though the uncertainties in semi-major axis (to directly compare with theoretical values in our calculations) of observed meteoroids from these highly successful observations are quite high (which is the typical case for all meteor observations, especially when the semi-major axis itself is high), the matching of outburst time frames and solar longitudes from these papers itself is a very effective way of comparing the orbital evolution of resonant meteoroids with real observations. Our results show that 1:6 resonant meteoroids ejected from the resonant
comet also caused enhanced activity from 1933-1938 which match old
observational records (Millman 1936, Lovell 1954). Most of these meteoroids
had positive ejection velocities which is more favorable for stronger
outbursts as discussed before.

Fig. 8 clearly shows that meteoroids with initial orbital periods
corresponding to 1:6 (around 71 years) and 2:13 (around 77 years) resonances
have their ascending nodes near the orbit of the Earth. Hence it can be
concluded that resonance mechanisms (specifically 1:6 and 2:13 MMR in this
case) aid these particles to come near the Earth at the present epoch while
the non-resonant ones precess away from the Earth's orbit. This is typical of
other ejection epochs (as shown in Table 1) as well.  The ascending node of
Halley during its last apparition (in 1986) was 1.8 AU. One could clearly see
that the number of particles trapped in 1:6 resonance is considerably larger
than the number trapped in 2:13 resonance. Moreover in Fig. 5 we notice that
particles having orbital periods of almost 5$P_{j}$(59.2 years),
6$P_{j}$(71.1 years) and 7$P_{j}$(83.0 years) come near the Earth's orbit
which agrees with earlier calculations done by Sato \& Watanabe (2007).

The typical ZHR for Orionids during non-outburst years is about 20 (Rendtel \& Betlem 1993, Rendtel 2008, IMO records). From the recorded previous observations it is seen that the ZHR is about 60 (Rendtel 2007, Kero et al. 2011, IMO records) due to 1:6 MMR during 2006-2010 and about 35 (Rendtel \& Betlem 1993) due to 2:13 MMR in 1993. Using these previous observations and flux, one could actually make a simplistic estimation of the mass delivered to Earth from the Orionid stream during these outburst years. According to the detailed work of Hughes and McBride 1989, the typical influx rate (at shower maximum perpendicular to the radiant) of Orionids is $1.8 \times 10^{-18} gcm^{-2}s^{-1}$  which in turn (after multiplying with the incident area of Earth) predicts 7 g/s during 2006-2010 and 4 g/s during 1993. It should be made clear that enhanced ZHR could be quite different in other outburst years (in past as well as future) as it depends on the exact cross section and density distribution of resonant trails intersecting the earth. Moreover, given the high speed of Orionid meteors and the strong dependence of meteor brightness on velocity, the meteoroidal mass influx to Earth is not dominated by this level (ZHR = a few tens) of Orionid outburst.

\begin{table*}
\centering
\caption{Data of dust trails which caused various Orionid outbursts}
\label{}
\begin{tabular}{@{}rccrrc}
\hline
\multicolumn{1}{c}{Ejection} &Expected peak time & $\lambda_\odot$ & $\Delta$r & \multicolumn{1}{c}{Ejection} & \multicolumn{1}{c}{Period at} \\
\multicolumn{1}{c}{year}     &    (UT)           & (J2000.0)       & (AU)      & velocity (m/s)               & ejection\ (years)    \\
\hline
-1197   & 1436\ Oct\ 13\ 01:44&207.551       &+0.0012                & +13.16             &71.76            \\
-985   & 1436\ Oct\ 14\ 17:40  & 208.212       &+0.0027               & -12.79   & 71.64        \\
-910   & 1437\ Oct\ 14\ 03:00 & 208.344          &-0.0013               &-13.60  &71.95          \\
-836  & 1438\ Oct\ 14\ 13:30  & 208.522           &+0.0010              & -22.08   &70.12         \\
-1265   & 1439\ Oct\ 14\ 23:54 & 208.698        &+0.0021              &+11.88      &  70.67          \\
-985   & 1439\ Oct\ 15\ 00:00  & 208.702           &-0.0001              & -14.74     &  71.23         \\
-910   & 1439\ Oct\ 16\ 15:03  & 210.324            &-0.0008             & -18.07     & 71.01          \\
-910   & 1440\ Oct\ 13\ 19:17  & 208.258             &+0.0037              & -20.37    & 70.54          \\[2mm]

-985   & 1916\ Oct\ 17\ 07:40& 204.771            &+0.0002        &+10.64      & 77.22        \\
-910   & 1916\ Oct\ 17\ 12:57& 204.990          &+0.0014        &+8.79      & 77.29        \\[2mm]

-1265  & 1933\ Oct\ 21\ 02:24& 208.170       &+0.0011          &+15.69    & 71.54     \\
-985  & 1933\ Oct\ 21\ 02:52& 208.190       &+0.0086          &-12.18    & 71.77     \\
-1333  & 1934\ Oct\ 21\ 12:14& 208.321       &+0.0044          &+9.13    & 71.82     \\
-985  & 1934\ Oct\ 21\ 12:28& 208.330       &+0.0002          &-11.16    & 71.99     \\
-985  & 1935\ Oct\ 21\ 13:26& 208.120       &+0.0034          &-16.42    & 71.88     \\
-910  & 1935\ Oct\ 22\ 05:16& 208.771       &+0.0087          &-14.35    & 71.79     \\
-1265  & 1936\ Oct\ 21\ 16:19& 208.982       &-0.0073          &+17.35    & 71.93     \\
-1197  & 1936\ Oct\ 22\ 06:28& 209.568       &+0.0058          &+10.77    & 71.21     \\
-1265  & 1937\ Oct\ 21\ 20:24& 208.890       &+0.0067          &+17.27    & 71.91     \\
-836  & 1937\ Oct\ 21\ 23:02& 208.998       &+0.0022          &-13.68    & 71.85     \\
-1197 & 1938\ Oct\ 21\ 21:21& 208.675       &-0.0091          &+7.00    & 71.37     \\
-1265  & 1938\ Oct\ 22\ 02:24& 208.883       &+0.0001          &+14.31    & 71.22     \\[2mm]

-1333   & 1993\ Oct\ 17\ 22:48&204.662   &+0.0039           &+31.88  &  77.70      \\
-985  & 1993\ Oct\ 18\ 00:14 & 204.724     &+0.0041           &+12.68  &  77.77        \\
-910   & 1993\ Oct\ 18\ 02:26 & 204.774     &-0.0017          &+8.98      & 77.34        \\
-835   & 1993\ Oct\ 18\ 02:40& 204.782       &+0.0088           &+10.45   &   77.64       \\[2mm]

-1265   & 2006\ Oct\ 21\ 02:09& 207.452      &+0.0005         &+11.03   &  70.48        \\
-910   & 2006\ Oct\ 23\ 03:38& 209.461      &-0.0069         &-17.73   &  71.08        \\
-1333   & 2007\ Oct\ 21\ 18:14& 207.858      &+0.0043         &+8.60   &  71.70        \\
-1265   & 2007\ Oct\ 22\ 00:28 & 208.118     &-0.0004          &+12.95   &  70.91        \\
-1197  & 2007\ Oct\ 22\ 04:36& 208.252       &+0.0032          &+11.25    & 71.32     \\
-985  & 2007\ Oct\ 22\ 09:21&208.486           &+0.0047         &-15.22   &   71.13       \\
-910  & 2007\ Oct\ 22\ 10:04&208.513            &+0.0002        &-15.81   &  71.48    \\
-836  & 2007\ Oct\ 23\ 01:12  & 209.140          &-0.0071        &-15.76   &   71.41       \\
-1333  & 2008\ Oct\ 23\ 05:16\ & 210.050       &+0.0073         &+6.61   &  71.25       \\
-1265  & 2008\ Oct\ 20\ 14:40\  & 207.412      &+0.0022         &+14.65  &  71.30        \\
-1197  & 2008\ Oct\ 21\ 07:28\   &208.114        &-0.0019         &+9.30    & 70.88     \\
-985  & 2009\ Oct\ 21\ 15:07\     &208.214         &+0.0077         &-18.72   &   70.41      \\
-910  & 2009\ Oct\ 21\ 19:43\    &208.367           &-0.0008        &-17.34    & 71.16     \\
-836  & 2010\ Oct\ 21\ 21:36\     & 208.225        &-0.0011        &-17.88   &   70.97       \\
-1333   & 2010\ Oct\ 21\ 23:31\  &208.302         &+0.0056          &+6.30   & 71.18      \\
-1265   & 2010\ Oct\ 22\ 02:24\  &208.424         &+0.0013         &+15.26      & 71.44           \\[2mm]

-910   & 2070\ Oct\ 18\ 19:12\  & 204.770          &-0.0029         & +9.24     & 77.41        \\
-910   & 2070\ Oct\ 18\ 19:26\ & 204.779            &+0.0003        &+9.80      & 77.56        \\

\hline
\end{tabular}\\
\end{table*}

\section{Orbit of Halley before 1404 B.C.}

Our calculations show that the orbit of Halley was substantially different
from the present orbit at about 12,000 years in the past. Fig. 9 shows the
time evolution of semi-major axis, indicating a drastic change in the
semi-major axis near this time frame. A similar sudden change occurred in
eccentricity, inclination and longitude of pericentre. Fig. 10 plots the time
evolution of heliocentric distance of descending node, showing that close
encounters with Jupiter are the reason for this drastic variation in the
comet's orbit.  100 clones with orbits very similar (varying semi-major axis
and eccentricity minutely while keeping the perihelion distance as constant)
to the comet were integrated 30,000 years backwards in time from 240 B.C.
and this behavior is typical for about 95\% of the clones.

From these orbital integrations it is clear that any meteoroid ejection
before 12,000 years in the past would not correspond to the present day
Orionid meteor shower. Hence this particular time constraint can be used as a
starting epoch for ejection to simulate the present day Orionid stream. It is
also interesting to note that this timescale is close to the physical
lifetime of the comet itself. In our test simulations, almost 80\% of the
clones get trapped into 1:6 and 2:13 resonances for at least a few thousand
years between -12,000 and -1403. Hence it is confirmed that the phenomenon of
resonance plays a vital role in the long term dynamical evolution of Halley
itself which further stresses the motivation in looking into more resonant
structures in the present day Orionid stream. This gives good scope for a lot
of interesting further work.

\section{Main Results}

We find that dust trails formed by 2:13 resonant meteoroids caused the
unusual meteor outbursts on 1993 Oct 18 (Miskotte 1993, Rendtel \& Betlem 1993) and 1916 Oct 17 (Olivier 1921).  Meteor outbursts
from 1436-1440 and 1933-1938 were due to the 1:6 resonance mechanism which
matches historical observations in 1436 and 1439 (Imoto \& Hasegawa 1958) and
1933-1938 (Millman 1936, Lovell 1954).  Furthermore we are able to correlate
the recent observations of outbursts from 2006-2010 (Rendtel 2007, Trigo-Rodriguez et al.\ 2007, Arlt et al.\ 2008, Kero et al.\ 2011, IMO
database) due to 1:6 resonant
meteoroids with our theoretical simulations. These correlations are very
promising and give us great confidence in confirming theory with observations. Using similar techniques one could also predict similar events for the future. We foresee a meteor outburst in
2070 (due to the 2:13 resonance) similar to the 1993 outburst. Using the data (ZHR and mass flow rate) from previous observations it is also possible to roughly estimate the mass influx in outburst years (see Section 3.2).

Although non-resonant particles can produce random outbursts, our calculations show
that a substantial majority of Orionid outbursts are due to resonant
structures in the meteor stream. However it must be pointed out that much older meteoroids (ejected
before 1404 B.C.) may also contribute to all these outbursts (which makes
further backward integrations and calculations very crucial). 

\section{Conclusions and Future Work}

Most of the theoretical aspects of these two resonances can have very significant and interesting effects on real observations.
The compact dust trails getting preserved for many 10 kyr due to 1:6 MMR hint at an
exciting possibility that strong meteor outbursts could occur in the future
even after the comet becomes extinct, i.e.\ survival times of some resonant
structures could be much higher than the physical lifetime of the parent
body.

It is well known and obvious that understanding the history of comets is crucial to
predicting meteor showers. In this work we find that Orionid outbursts in 1436-1440,1916,1933-1938, 1993 and 2006-2010 were caused by resonant particles ejected
from Halley before 240 B.C., the date beyond which there are no direct
observational records of the comet. As a corollary to the above point about the importance of knowing comets' histories, one could argue that non-uniform meteor rates can act as a
great tool to backtrack the history of a comet beyond the time frame in which there are direct
sightings of the comet itself. All of these prove how useful the comparison between meteor observations and these simulations are. In short it is an indirect confirmed observation of the comet beyond 240 B.C. 

Even though the Eta Aquariid
shower is considerably different (McIntosh \& Hajduk 1983) from the Orionid
shower in many ways, it would be worthwhile to verify whether all these
resonant phenomena and enhanced activity are applicable in its case as
well (CBET 944, 2007). The low number (compared to Orionids) of credible observations of
Eta-Aquariids is a limitation in this regard though. Near future releases
 of radio results on Eta Aquariids (personal communications with Campbell-Brown and Jenniskens) would be very promising in this direction. Negative
observations, comprising diminished meteor rates of Orionids in some particular years
compared to adjacent years (e.g.\ ZHR reaching only 7 in 1900: Kronk 1988)
could be as scientifically valuable as enhanced meteor phenomena which we have investigated in this work. A future careful study of such events can also be intriguing in many aspects.

 As the next step we plan to design an ejection model to simulate the Orionid
stream beginning 12,000 years in the past and to correlate more past and
present observations as accurately as possible. 

\section{Acknowledgements}

The authors wish to thank both the anonymous reviewers for the helpful comments and also intend to express their gratitude to the Department of Culture, Arts
and Leisure of Northern Ireland for the generous funding to pursue astronomical research at Armagh Observatory.

\section{References}

Arlt R., Rendtel J. and Bader P.  2008. The 2007 Orionids from visual observations.  WGN (Journal of the International Meteor Organisation) 36: 55-60\

Asher D. J. and Emel'yanenko V. V. 2002. The origin of the June Bootid outburst in 1998 and determination of cometary ejection velocities. Monthly Notices of the Royal Astronomical Society 331: 126-132\

Asher D. J., Bailey M. E. and Emel'yanenko V. V. 1999. Resonant meteoroids from Comet Tempel-Tuttle in 1333: the cause of the unexpected Leonid outburst in 1998. Monthly Notices of the Royal Astronomical Society 304: L53-56\

Bailey M. E. and Emel'yanenko V. V.  1996. Dynamical evolution of Halley-type comets. Monthly Notices of the Royal Astronomical Society 278: 1087-1110\

Central Bureau Electronic Telegram No. 944 dated 2007 Apr 25\\ (http://www.cbat.eps.harvard.edu/iau/cbet/000900/CBET000944.txt)\

Chambers J. E. 1999. A hybrid symplectic integrator that permits close encounters between massive bodies. Monthly Notices of the Royal Astronomical Society 304: 793-799\

Christou A. A., Vaubaillon J., and Withers P. 2008. The P/Halley Stream: Meteor Showers on Earth, Venus and Mars. Earth, Moon, and Planets, 102: 125-131\

Denning W. F. 1899. Meteoric showers in autumn, winter, and spring from Ursa Major and the region near. The Observatory 22: 90-91\

Dubietis A. 2003. Long-term activity of meteor showers from Comet 1P/Halley. WGN (Journal of the International Meteor Organisation) 31: 43-48 \

Emel'yanenko V. V. 1988. Meteor-stream motion near commensurabilities with Jupiter. Soviet Astronomy Letters 14: 278-281.

Emel'yanenko V. V. 2001. Resonance structure of meteoroid streams. In Proceedings of the Meteoroids 2001 Conference (ESA SP--495), Edited by Warmbein B. Noordwijk:  ESA. pp.\ 43-45.\

Everhart E. 1985. An efficient integrator that uses Gauss-Radau spacings. In Proceedings of IAU Colloq. 83, Edited by Andrea Carusi and Giovanni B. Valsecchi. Astrophysics and Space Science Library 115: 185-202\

Giorgini J.D., Yeomans D.K., Chamberlin A.B., Chodas P.W., Jacobson R.A., Keesey M.S., Lieske J.H., Ostro S.J., Standish E.M., and Wimberly R.N. 1996. JPL's On-Line Solar System Data Service, Bulletin of the American Astronomical Society 28(3), 1158\

Hajduk A. 1986. Meteoroids from Comet Halley and the comets mass production and age. In ESA Proceedings of the 20th ESLAB Symposium on the Exploration of Halley's Comet, 2: 239-243\

Halley E. 1705. Synopsis Astronomiae Cometicae. London\

Herschel A. S. 1866. Radiant points of shooting stars. Monthly Notices of the Royal Astronomical Society 26: 51-53\
 
Hughes D. W. 1985. The size, mass, mass loss and age of Halley's comet. Monthly Notices of the Royal Astronomical Society 213, 103-109\

Hughes D.W. and McBride N. 1989. The mass of meteoroid streams. Monthly Notices of the Royal Astronomical Society 240: 73-79\ 

Imoto S. and Hasegawa I. 1958. Historical Records of Meteor Showers in China, Korea, and Japan. Smithsonian Contributions to Astrophysics 2: 131-144\

International Meteor Organisation Records (http://www.imo.net/zhr)\

Jenniskens P. 2006. Meteor showers and their parent comets.   Cambridge, UK:  Cambridge University Press.\

Jenniskens P., Lyytinen E., Nissinen M., Yrjola I., and Vaubaillon J. 2007.
Strong Ursid shower predicted for 2007 December 22.
WGN (Journal of the International Meteor Organisation) 35: 125-133.\

Kepler J. 1609. Astronomia Nova. Heidelberg\

Kepler J. 1619. Harmonices Mundi Libri V. Linz.\

Kero J., Szasz C., Nakamura T., Meisel D.D., Ueda M., Fujiwara Y., Terasawa T., Miyamoto H. and Nishimura K. 2011. First results from the 2009-2010 MU radar head echo observation programme for sporadic and shower meteors: the Orionids 2009. Monthly Notices of the Royal Astronomical Society 416: 2550-2559\ 

Kondrat'eva E.D. and Reznikov E.A. 1985. Comet Tempel-Tuttle and the Leonid meteor swarm. Solar System Research 19: 96-101.\

Kozai Y. 1979. Secular perturbations of asteroids and comets. In Dynamics of the solar system. Proceedings IAU Symposium 81. Edited by Duncombe R.L.  Dordrecht: Reidel Publishing Co., pp. 231-236\

Kresak L. 1987. The Evolution of the Small Bodies of the Solar System. Proceedings of the International School of Physics "Enrico Fermi". Edited by M. Fulchignoni, and L. Kresak.  Amsterdam:  Elsevier. pp.\ 202-216

Kronk G. W. 1988.  Meteor Showers: a descriptive catalog.  Hillside, NJ:  Enslow Publishers.

Lindblad B. A. and Porubcan V. 1999. Orionid Meteor Stream. Contributions of the
 Astronomical Observatory of Skalnat\'e Pleso 29: 77-88\

Lovell A. C. B. 1954, Meteor Astronomy.  Oxford: Clarendon Press.\

Marsden B. G. and Williams G. V. 2008. Catalogue of Cometary Orbits, 17th ed.  Cambridge, MA:  Minor Planet Center/Central Bureau for Astronomical Telegrams.

Marsden B.G., Sekanina Z. and Yeomans D.K. 1973. Comets and non-gravitational forces. V. The Astronomical Journal 78: 211-225\

McIntosh B. A. and Hajduk A. 1983. Comet Halley meteor stream - A new model. Monthly Notices of the Royal Astronomical Society 205: 931-943\

McIntosh B. A. and Jones J. 1988. The Halley comet meteor stream- Numerical modelling of its dynamic evolution. Monthly Notices of the Royal Astronomical Society 235: 673-693\

Millman P. M. 1936. Meteor News (Observation of the Orionids in 1936). Journal of the Royal Astronomical Society of Canada 30: 416-418 \

Miskotte K. 1993. High Orionid activity on October 18, 1993. WGN (Journal of the International Meteor Organisation) 21: 292\ 

Murray C.D. and Dermott S.F. 1999. Solar system dynamics.  Cambridge, UK:  Cambridge University Press.\

Newton I. 1687. Philosophiae Naturalis Principia Mathematica. London.\

Olivier C.P. 1921. Parabolic orbits of meteor streams. Publications of the Leander McCormick Observatory 2: 201-268\

Rendtel J. 2007. Three days of enhanced Orionid activity in 2006 - Meteoroids from a resonance region?. WGN (Journal of the International Meteor Organisation) 35: 41-45\

Rendtel J. 2008. The Orionid Meteor Shower Observed Over 70 Years. Earth Moon, and Planets, 102: 103-110\

Rendtel J. and Betlem H. 1993. Orionid meteor activity on October 18, 1993. WGN (Journal of the International Meteor Organisation) 21: 264-268\

Ryabova G. 2003. The comet Halley meteoroid stream: just one more model. Monthly Notices of the Royal Astronomical Society 341: 739-746\

Ryabova G. O. 2006. Meteoroid streams: mathematical modelling and observations. In Asteroids, Comets, Meteors, Proceedings\ IAU Symposium\ 229, edited by Lazzaro D., Ferraz-Mello S., Fern\'andez J. A.  Cambridge, UK: Cambridge University Press. pp.\ 229-247.\

Saha P. and Tremaine S. 1993. The orbits of the retrograde Jovian satellites. Icarus 106: 549-562\

Sato M. and Watanabe J. 2007. Origin of the 2006 Orionid Outburst. Publications of the Astronomical Society of Japan 59: L21-L24\

Soja R. H., Baggaley W. J., Brown P., and Hamilton D. P. 2011.  Dynamical resonant structures in meteoroid stream orbits.  Monthly Notices of the Royal Astronomical Society 414: 1059--1076.\

Spurny P. and Shrbeny L. 2008. Exceptional Fireball Activity of Orionids in 2006. Earth, Moon, and Planets 102: 141-150 \

Steel D. I. 1987. The dynamical lifetime of comet P/Halley. Astronomy and Astrophysics 187: 909-912\

Trigo-Rodriguez J.M., Madiedo J.M., Llorca J., Gural P.S., Pujols P., Tezel T. 2007. The 2006 Orionid outburst imaged by all-sky CCD cameras from Spain: meteoroid spatial fluxes and orbital elements. Monthly Notices of the Royal Astronomical Society 380: 126-132\ 

Vaubaillon J., Lamy, P., and Jorda, L. 2006.
On the mechanisms leading to orphan meteoroid streams.
Monthly Notices of the Royal Astronomical Society 370: 1841-1848.\

Whipple F. L. 1951. A Comet Model. II. Physical Relations for Comets and Meteors. The Astrophysical Journal 113: 464-474\

Whipple A. L. and Shelus P. J. 1993. A secular resonance between Jupiter and its eighth satellite?. Icarus 101: 265-271\

Williams I. P. 2011. The origin and evolution of meteor showers and meteoroid streams. Astronomy and Geophysics 52: 2.20-2.26\

Yeomans D. K. and Kiang T. 1981. The long-term motion of comet Halley.  Monthly Notices of the Royal Astronomical Society 197: 633-646\

Zhuang T. S. 1977. Ancient Chinese reports of meteor showers. Chinese Astronomy 1: 197-220\

\end{document}